\documentclass[conference]{IEEEtran}
\IEEEoverridecommandlockouts

\usepackage{cite}
\usepackage{amsmath,amssymb,amsfonts}
\usepackage{algorithmic}
\usepackage{graphicx}
\usepackage{textcomp}
\usepackage{xcolor}
\usepackage{bigstrut}
\usepackage{array}
\usepackage{epsfig}
\usepackage{multirow}
\def\BibTeX{{\rm B\kern-.05em{\sc i\kern-.025em b}\kern-.08em
    T\kern-.1667em\lower.7ex\hbox{E}\kern-.125emX}}
\begin{document}

\title{DULRTC-RME: A Deep Unrolled Low-rank Tensor Completion Network for Radio Map Estimation
\thanks{This work was supported in part by the National Natural Science Foundation of China under Grants U2066201 and 62101054, in part by the Fundamental Research Funds for the Central Universities under Grant No. 24820232023YQTD01, and in part by the Double First-Class Interdisciplinary Team Project Funds 2023SYLTD06 (Corresponding author: Li Wang, Email: liwang@bupt.edu.cn).}
}

\author{
\IEEEauthorblockN{Yao Wang\textsuperscript{${\star}$}, Xin Wu\textsuperscript{${\star}$}, Lianming Xu\textsuperscript{${\dagger}$}, Na Liu\textsuperscript{${\star}$} and Li Wang\textsuperscript{${\star}$}}
\IEEEauthorblockA{\textsuperscript{${\star}$}School of Computer Science (National Pilot Software Engineering School), \\Beijing University of Posts and Telecommunications, Beijing 100876, China \\
\textsuperscript{${\dagger}$}School of Electronic Engineering, Beijing University of Posts and Telecommunications, Beijing 100876, China}
}

\maketitle

\begin{abstract}
Radio maps enrich radio propagation and spectrum occupancy information, which provides fundamental support for the operation and optimization of wireless communication systems. Traditional radio maps are mainly achieved by extensive manual channel measurements, which is time-consuming and inefficient. To reduce the complexity of channel measurements, radio map estimation (RME) through novel artificial intelligence techniques has emerged to attain higher resolution radio maps from sparse measurements or few observations. However, black box problems and strong dependency on training data make learning-based methods less explainable, while model-based methods offer strong theoretical grounding but perform inferior to the learning-based methods. In this paper, we develop a deep unrolled low-rank tensor completion network (DULRTC-RME) for radio map estimation, which integrates theoretical interpretability and learning ability by unrolling the tedious low-rank tensor completion optimization into a deep network. It is the first time that algorithm unrolling technology has been used in the RME field. Experimental results demonstrate that DULRTC-RME outperforms existing RME methods.
\end{abstract}


\section{Introduction}
\label{sec:intro}
As a fundamental tool for wireless communication systems, radio maps generally reveal important information in multiple domains (e.g., space and frequency) of the communication environment, describing communication environments through channel modelling parameters, such as the geometric signal power spectral density (PSD) \cite{b1}. However, dense and complete radio maps are not directly available and generally tend to be estimated from sparse observations manually collected by sensors in the target area. The observations may be sparse in both spatial and frequency domains. Therefore, a main task called radio map estimation (RME) is to recover relatively higher resolution radio maps from them \cite{b2}.

Early attempts for RME can be broadly classified into two categories: model-based and learning-based methods. Model-based methods generally assume specific radio propagation models. For instance, \cite{b3} uses the log-distance path loss (LDPL) interpolation model to estimate WiFi radio maps. 
However, these theoretical models often struggle to fully adapt to complex real-world scenarios, as they fail to capture intricate structures like shadows or obstacles, leading to limited accuracy in RME.

Instead of relying on pre-specified, hand-crafted model priority, learning-based methods ``learn" from sparse observations and utilize deep neural networks (DNNs) to capture complex underlying structures in radio data \cite{b5,b6,b7}. Although showing promise in handling complex radio environments, learning-based methods heavily depend on the quantity and quality of training samples, which are often limited and biased in real-world scenarios. Besides, DNN architectures are typically designed empirically, making their behaviour and performance difficult to analyze and predict.

\begin{figure}[t!]
\centerline{\includegraphics[height=6.5cm,width=8cm]{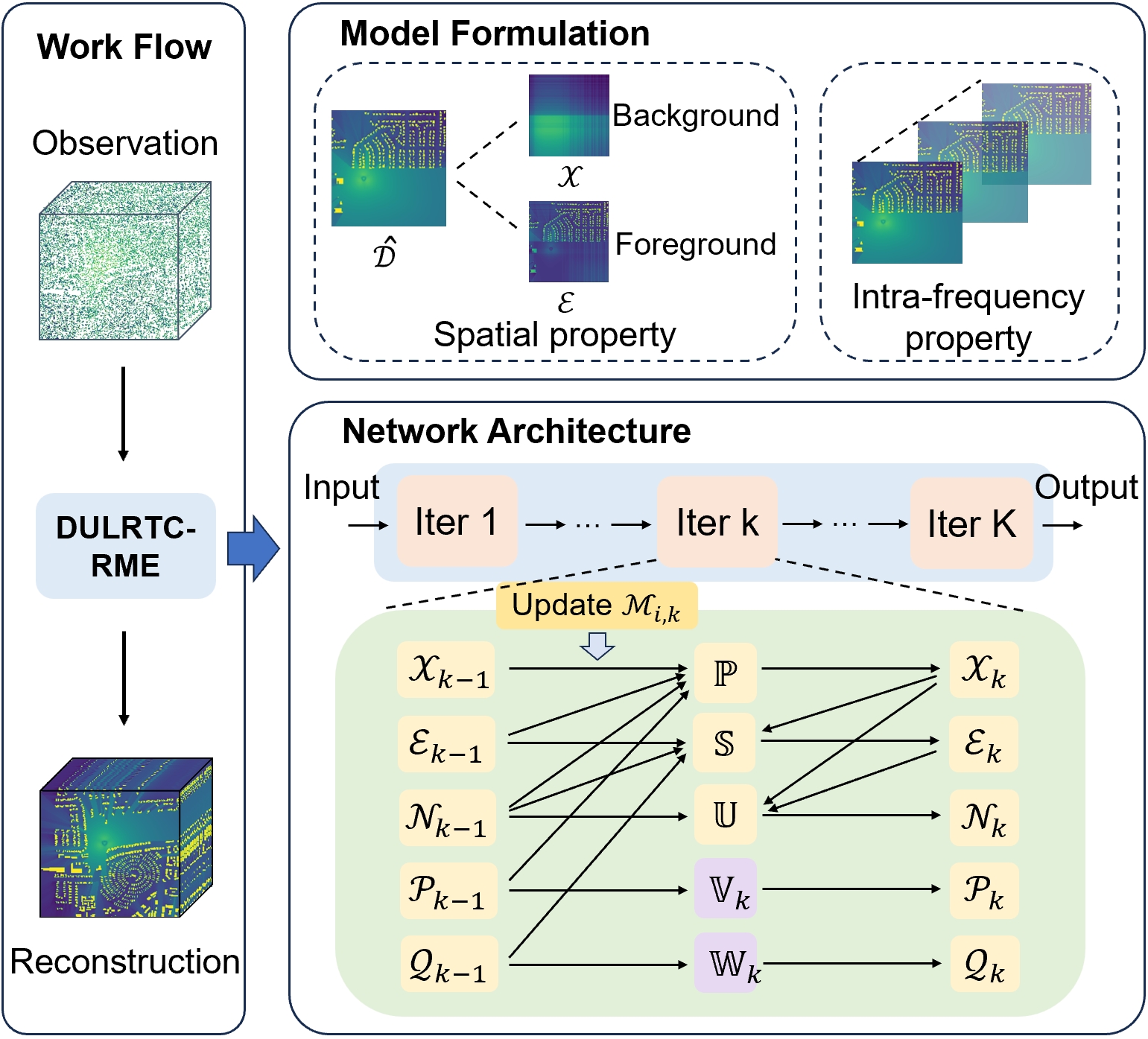}}
\vspace{-0.5em}
\caption{The flowchart of the proposed DULRTC-RME.}
\label{fig:archichart}
\end{figure}

To leverage the merits of both model-based and learning-based methods while overcoming their drawbacks, the idea of physics-inspired machine learning methods \cite{b2}  has been introduced to improve RME. A notable method among them called algorithm unrolling \cite{b9}  bridges deep networks and traditional models by implementing each iteration of the optimization algorithm as a layer of the DNN. Inspired by this, a deep unrolled network based on low-rank tensor completion (DULRTC) for RME is proposed in this work, as shown in Fig.~\ref{fig:archichart}.  Specifically, we model RME as a low-rank tensor completion problem by considering radio propagation fading characteristics and introducing regularization functions to account for inaccuracies in the decomposition assumption. The optimization problem is solved iteratively. By decomposing the iterative algorithm into modules, we transform it into the equivalent DNN, which maintains interpretability owing to its theoretical foundation in low-rank tensor completion. Despite the wide application of algorithm unrolling in image and video processing, to our knowledge, there has been limited research on RME. Experiments on the BART-Lab Radiomap Dataset demonstrate that DULRTC-RME outperforms traditional interpolation methods and purely deep learning-based approaches in recovering more accurate and higher-quality radio maps.

\section{Proposed DULRTC-RME Algorithm}
\label{sec:proposed}
\subsection{Mathematical Preliminaries}
\label{ssec:preliminaries}
In this paper, scalars, vectors, matrices and tensors are denoted respectively by italic lower case letters (eg., \textbf{x}), bold lower case letters (eg., $\boldsymbol{x}$), italic upper case letters (eg., $\textit{X}$), and bold calligraphic letters (eg., $\mathcal{X}$). 
The element in a $K$th-order tensor $\mathcal{X} \in R^ {I_1 \times I_2 ... \times ... I_K}$ at location $(i_1,i_2,...i_K)$ is denoted by $\mathcal{X}(i_1,i_2,...i_K)$. 
For a third-order tensor $\mathcal{X}\in\mathbb{R}^{n_1\times{n_2}\times{n_3}}$, we can present its $i$th horizontal, lateral, and frontal slices as $\mathcal{X}(i,:,:)$, $\mathcal{X}(:,i,:)$, $\mathcal{X}(:,:,i)$. 
The inner product of two tensors $\mathcal{X},\mathcal{Y}\in\mathbb{R}^{n_1\times{n_2}\times...\times{n_K}}$ is defined as $<\mathcal{X},\mathcal{Y}>=\sum_{i_1,i_2,...,i_K}\mathcal{X}(i_1,i_2,...,i_K)\mathcal{Y}(i_1,i_2,...,i_K)$. 

If we retain only one degree of freedom in the tensor while fixing all other dimensions, we obtain fibers in the form of strips. To compute the mode-$m$ unfolding of a tensor, we extract fibers according to different modes and rearrange them into combinations. The mode-$m$ unfolding of a tensor $\mathcal{X}$ is represented as $\mathcal{X}_{(m)}\in\mathbb{R}^{n_k\times{I_m}}$, where $I_m=\prod_{k=1,k\neq m}^N n_k$. The tensor element $(i_1,i_2,...i_K)$ maps to the matrix element $(i_m, j)$, with $j=1+\sum_{k-1,k\neq m}^N(i_k-1)J_k$ and $J_k=\prod_{l=1,l\neq m}^{k-1}n_l$.


\subsection{Problem Formulation}
\label{ssec:problem}
Assume the target geographical region (with a grid size of $H\times{W}$) contains several transmitters operating in specific $K$ frequency bands, the sparse measurements are represented as a third-order tensor $\mathcal{D} \in{\mathbb{R}^{H\times{W} \times{K}}}$. Given that sensors collect PSD values at specific locations $\Omega$, and the reconstructed radio map is $\hat{\mathcal{D}}\in{R^{H\times{W}\times{K}}}$, we have $\mathcal{P}_\Omega(\mathcal{D}) = \mathcal{P}_\Omega(\hat{\mathcal{D}})$.

Based on the complex fading characteristics of radio signal propagation, the desired radio map $\hat{\mathcal{D}}$ can be decomposed into two components: a uniform background $\mathcal{X}$ and a sparse foreground $\mathcal{E}$. In the spatial domain, large-scale fading occurs as signals propagate over long distances between transceivers, resulting in path loss that is primarily dependent on the transmission path, creating a regular, uniform pattern. In the frequency domain, radio maps at different frequencies within the same spatial scenario are linearly correlated. Thus, $\mathcal{X}$ exhibits low rank in both spatial and frequency domains. Meanwhile, $\mathcal{E}$ remains sparse because small-scale fading, resulting from reflection, scattering, and diffraction, combined with geospatial architectural features, causes most values in the radio map to be zero. Therefore, a multi-frequency RME problem is formulated as
\begin{equation}
\setlength{\abovedisplayskip}{2pt}
\setlength{\belowdisplayskip}{3pt}
\begin{aligned}
\mathop{\min}_{\mathcal{X},\mathcal{E}}& \quad \sum_{i=1}^{3} \alpha_i\Vert \mathcal{X}_{(i)} \Vert_*+\lambda \Vert\mathcal{E}\Vert_1,
\\& \text{s.t.,}\quad \mathcal{P}_\Omega(\mathcal{X}+\mathcal{E})= \mathcal{P}_\Omega(\mathcal{D}).
\end{aligned}
\end{equation}

We use the sum of the nuclear norm of the tensor unfold in different modes as a convex approximation of the tensor rank since it is challenging to solve directly. The above model based on certain assumptions may not accurately represent all features of a real-world scenario. To tackle the limitations, we follow the approach in \cite{b13} and add regularisation functions $f:{ \mathbb{R}^{H\times{W}\times{K}}}\rightarrow \mathbb{R}$ and $g:{ \mathbb{R}^{H\times{W}\times{K}}}\rightarrow  \mathbb{R}$ to compensate for modelling inaccuracy. In addition, to account for noise in the real case, we introduce relaxation variable $\mathcal{N}$ and ultimately reformulate the optimization problem as
\begin{equation}
\setlength{\abovedisplayskip}{2pt}
\setlength{\belowdisplayskip}{3pt}
\begin{aligned}
&\mathop{\min}_{\mathcal{X},\mathcal{E}}\quad\sum_{i=1}^{3}\alpha_i\Vert\mathcal{X}_{(i)}\Vert_*+\lambda\Vert\mathcal{E}\Vert_1 +f(\mathcal{X})+g(\mathcal{E}), 
\\& \text{s.t.,}\;\mathcal{X}+\mathcal{E}+\mathcal{N}= \mathcal{P}_\Omega(\mathcal{D})\,~\text{and}~\,\Vert\mathcal{P}_\Omega(\mathcal{N})\Vert_F\leq\delta, 
\label{eq:formulation}
\end{aligned}
\end{equation}
where $\delta$ is the noise threshold.

\vspace{-3mm}
\subsection{Solution to the Optimization}
\label{ssec:solution}
Eq.~\eqref{eq:formulation} is a convex problem with non-differentiable terms and interrelated parameters. Thus, we introduce auxiliary variables $\mathcal{P}$ and $\mathcal{Q}$ and adopt the multiplier alternating direction method (ADMM) \cite{b14} to split the optimization problem into $\mathcal{X},\mathcal{E},\mathcal{N},\mathcal{P},\mathcal{Q}$-subproblems separately. The augmented Lagrangian function is then formulated as
\vspace{-3mm}
\begin{align}
\setlength{\abovedisplayskip}{3pt}
\setlength{\belowdisplayskip}{0pt}
    &\mathcal{L}(\mathcal{X},\mathcal{E},\mathcal{N},\mathcal{P},\mathcal{Q},\Lambda,\Gamma,\Phi)=\nonumber
    \sum_{i=1}^{3}\alpha_i\Vert\mathcal{X}_{(i)}\Vert_*+\lambda\Vert\mathcal{E}\Vert_1+
    \\&<\Lambda,{\mathcal{P}_\Omega(\mathcal{D})-\mathcal{X}-\mathcal{E}-\mathcal{N}}>\nonumber+
    \frac{\mu}{2}\Vert{\mathcal{P}_\Omega(\mathcal{D})-\mathcal{X}-\mathcal{E}-\mathcal{N}}\Vert_F^2\nonumber
    \\&+f(\mathcal{P})+<\Gamma,{\mathcal{X}-\mathcal{P}}>+\frac{\theta}{2}\Vert{\mathcal{X}-\mathcal{P}}\Vert_F^2\nonumber
    \\&+g(\mathcal{Q})+<\Phi,{\mathcal{E}-\mathcal{Q}}>+\frac{\beta}{2}\Vert{\mathcal{E}-\mathcal{Q}}\Vert_F^2, 
\end{align}
where $\mu,\theta,\beta>0$, the penalty term $\Lambda,\Gamma,\Phi \in\mathbb{R}^{H\times W\times K}$ is the Lagrange multiplier tensor. 

\textbf{$\mathcal{X}$-subproblem:} By taking the partial derivative of $\mathcal{L}$ to $\mathcal{X}$,  $\mathcal X$ is updated by
\begin{small}
\begin{equation}
\setlength{\abovedisplayskip}{3pt}
\setlength{\belowdisplayskip}{3pt}
\begin{split}
\mathcal{X}_{k+1}\!=\!&\mathop{\arg\min}\limits_{\mathcal{X}}\;\frac{1}{\mu_k\!+\! \theta_k}\sum_{i=1}^{3}\alpha_{i,k}\Vert\mathcal{X}_{(i)}\Vert_*
\!+\! \frac{1}{2}\Vert \mathcal{X}\!-\! \Psi_{\mathcal{X},k}\Vert_F^2,
\end{split}
\end{equation}
\end{small}
where $\Psi_{\mathcal{X},k}=\frac{1}{\mu_k+\theta_k}(\Lambda_k+\mu_k\mathcal{P}_\Omega(\mathcal{D})-\mu_k\mathcal{E}_k-\mu_k\mathcal{N}_k+\theta_k\mathcal{P}_k-\Gamma_k)$. Because ${\mathcal X}_{(i)}$s are interdependent when computing the nuclear norm, new auxiliary parameters $\mathcal{M}_i=\mathcal{X},\; {i=1,2,3}$ and corresponding Lagrange multipliers $\mathcal{Y}_i$s are introduced. The problem is split into $\mathcal{M}_i$-subproblem and $\mathcal{X}$-subproblem respectively. The solution below refers to the HaLRTC algorithm in \cite{b10},
\begin{equation}
\setlength{\abovedisplayskip}{3pt}
\setlength{\belowdisplayskip}{0pt}
\mathcal{M}_{i,k+1}=\text{fold}[D_\frac{\alpha_i}{\rho}(\mathcal{X}_{k,(i)}+\frac{1}{\rho}(\mathcal{Y}_{i,k})_{(i)})],
\end{equation}
\begin{equation}
\setlength{\abovedisplayskip}{3pt}
\setlength{\belowdisplayskip}{0pt}
\begin{split}
\mathcal{X}_{k+1}&=\frac{\rho\sum_{i=1}^3(\mathcal{M}_{i,k+1}-\frac{1}{\rho}\mathcal{Y}_{i,k})+(\mu_k+\theta_k)\Psi_{\mathcal{X},k}}{3\rho+\mu_k+\theta_k}.
\end{split}
\end{equation}

\textbf{$\mathcal{E}$-subproblem:} Next, we estimate $\mathcal{E}$ by solving the following problem
\begin{equation}
\setlength{\abovedisplayskip}{4pt}
\setlength{\belowdisplayskip}{3pt}
\begin{split}
\mathcal{E}_{k+1}&
=\mathop{\arg\min}\limits_{\mathcal{E}}\;\frac{\lambda_k}{\mu_k+\beta_k}\Vert\mathcal{E}\Vert_1+\frac{1}{2}\Vert\mathcal{E}-\Psi_{\mathcal{E},k}\Vert_F^2,
\end{split}
\end{equation}
where $\Psi_{\mathcal{E},k}=\frac{1}{\mu_k+\beta_k}(\Lambda_k+\mu_k\mathcal{P}_{\Omega}(\mathcal{D})-\mu_k\mathcal{X}_{k+1}-\mu_k\mathcal{N}_k+\beta_k\mathcal{Q}_k-\Phi_k)$. After applying the SVT algorithm, it turns to be 
\begin{equation}
\setlength{\abovedisplayskip}{3pt}
\setlength{\belowdisplayskip}{3pt}
\begin{split}
\mathcal{E}_{k+1}&=D_{\frac{\lambda_k}{\mu_k+\beta_k}}(\Psi_{\mathcal{E},k}).
\end{split}
\end{equation}

\textbf{$\mathcal{N}$-subproblem:} Set $lim_\mathcal{N} = \Vert\mathcal{P}_{\Omega}(\mathcal{N})\Vert_F\leq\delta_k$, we can estimate $\mathcal{N}$ by solving the below equation
\begin{equation}
\setlength{\abovedisplayskip}{3pt}
\setlength{\belowdisplayskip}{3pt}
\begin{split}
\mathcal{N}_{k+1}=&
\mathop{\arg\min}\limits_{\lim_\mathcal{N}}\;\Vert{\mathcal{N}-\Psi_{\mathcal{N},k}}\Vert_F^2,
\end{split}
\end{equation}
where $\Psi_{\mathcal{N},k}=\mathcal{P}_\Omega(\mathcal{D})-\mathcal{X}_{k+1}-\mathcal{E}_{k+1}+\frac{1}{\mu_k}\Lambda_k$.
It can be solved by \cite{b11}
\vspace{-1mm}
\begin{align}
    \mathcal{N}_{k+1}&=\mathcal{P}_{\Omega^C}(\Psi_{\mathcal{N},k})+\mathop{\min}\{\frac{\delta_k}{\Vert\mathcal{P}_{\Omega}(\Psi_{\mathcal{N},k})\Vert_F},1\}\mathcal{P}_{\Omega}(\Psi_{\mathcal{N},k}).
\end{align}

\textbf{$\mathcal{P}$ and $\mathcal{Q}$-subproblem:} The parameters $\mathcal{P}$ and $\mathcal{Q}$ in $k+1$th iteration can be express as:
\vspace{-3mm}
\begin{align}
    \mathcal{P}_{k+1}&=\mathop{\arg\min}\limits_{\mathcal{P}}\;f(\mathcal{P})+\frac{\theta_k}{2}\Vert\mathcal{P}-(\mathcal{X}_{k+1}+\frac{1}{\theta_k}\Gamma_k)\Vert_F^2\nonumber
    \\&=\text{prox}_f(\mathcal{X}_{k+1}+\frac{1}{\theta_k}\Gamma_k),
\end{align}
\vspace{-3mm}
\begin{align}
    \mathcal{Q}_{k+1}&=\mathop{\arg\min}\limits_{\mathcal{Q}}\;g(\mathcal{Q})+\frac{\beta_k}{2}\Vert\mathcal{Q}-(\mathcal{E}_{k+1}+\frac{1}{\beta_k}\Phi_k)\Vert_F^2\nonumber
    \\&=\text{prox}_g(\mathcal{E}_{k+1}+\frac{1}{\beta_k}\Phi_k),
\end{align}

where $\text{prox}_f(\cdot)$ and $\text{prox}_g(\cdot)$ denote the proximal operators of the regularisation functions $f(\cdot)$ and $g(\cdot)$, respectively. 

\subsection{Deep Unrolled Network}
We convert the iterative tensor completion algorithm into a series of blocks within a deep network. The ``Network Architecture" section in Fig.~\ref{fig:archichart} illustrates the structure of DULRTC-RME. The observation tensor is initially input, and all other variables are set to 0. The network consists of $K$ unrolling blocks, where optimization variables are iteratively updated. In each block, $\mathcal{X}$, $\mathcal{E}$, and $\mathcal{N}$ are updated via closed-form solutions. Specifically, we have $\mathcal{X}_{k+1}=\mathbb{P}(\mathcal{M}_{i,k+1}, \Psi_{\mathcal{X},k})$, $\mathcal{E}_{k+1}
=\mathbb{S}(\Psi_{\mathcal{E},k})$ and $ {\mathcal{N}_{k+1}}=\mathbb{U}(\Psi_{\mathcal{N},k})$. 

While for $\mathcal{P}_{k+1}$ and $\mathcal{Q}_{k+1}$, the regularisation function is traditionally determined according to certain applications, which may not be accurate enough and limits the model generalization. To address the constraints of hand-crafted regularizers, CNNs are introduced under the algorithm unrolling framework to implement the proximal operators, which can learn from the training data and effectively reconstruct complex and diverse visual features. We denote the CNNs in the $k$th iteration as $\mathbb{V}_k$ and $\mathbb{W}_k$, the solution is obtained by
\begin{equation}
\setlength{\abovedisplayskip}{3pt}
\setlength{\belowdisplayskip}{0pt}
\mathcal{P}_{k+1}=\mathbb{V}_k(\mathcal{X}_{k+1}+\frac{1}{\theta_k}\Gamma_k),    
\end{equation}
\begin{equation}
\setlength{\abovedisplayskip}{3pt}
\setlength{\belowdisplayskip}{0pt}
\mathcal{Q}_{k+1}=\mathbb{W}_k(\mathcal{E}_{k+1}+\frac{1}{\beta_k}\Phi_k).
\end{equation}

During training, the network parameters are updated based on input tensors $\mathbb{V}_k$ and $\mathbb{W}_k$ to obtain optimal solutions $\mathcal{P}_{k+1}$ and $\mathcal{Q}_{k+1}$.

The final output of the deep unrolling network is the estimated background $\mathcal{X}$ and foreground $\mathcal{E}$, which are merged to produce the desired result $\hat{\mathcal{D}}$.

In addition to the network parameters of  $\mathbb{V}_k$ and $\mathbb{W}_k$, the parameter $\mu_k$, $\theta_k$, $\beta_k$, $\lambda_k$ and $\delta_k$ are learnable and updated via back-propagation during training. This structure strictly follows the derived mathematical model, ensuring full interpretability and analytical capability. Since $\mathbb{V}_k$ and $\mathbb{W}_k$ are structure-agnostic nonlinear mappings, any CNN architecture can be used for them.

Finally, the Lagrange multiplier tensor is updated according to the ALM strategy
\begin{align}
    &\Lambda_{k+1}=\Lambda_{k}+\mu_k(\mathcal{P}_\Omega(\mathcal{D})-\mathcal{X}_{k+1}-\mathcal{E}_{k+1}-\mathcal{N}_{k+1}),
    \\&\Gamma_{k+1}=\Gamma_{k}+\theta_k(\mathcal{X}_{k+1}-\mathcal{P}_{k+1}),
    \\&\Phi_{k+1}=\Phi_{k}+\beta_k(\mathcal{E}_{k+1}-\mathcal{Q}_{k+1}),
    \\&\mathcal{Y}_{i,k+1}=\mathcal{Y}_{i,k}+\rho(\mathcal{X}_{k+1}-\mathcal{M}_{i,k+1}).
\end{align}

After introducing the architecture of the proposed network, we define the loss function as a combination of the baseline loss $L_{recon}$ and physical information loss $L_{phy}$, with $\omega$ controlling their contributions $L_{total}=\omega L_{recon}+(1-\omega)L_{phy}$. $L_{recon}$ is the $l_1$-norm between the estimated radio map and the ground truth. The physical information loss $L_{phy}$ \cite{b8} is defined as the mean square error (MSE) between the estimated and interpolated radio maps based on the radio propagation model LDPL.

\section{Experimental Results}
\newlength{\imagewidth}
\setlength{\imagewidth}{2.8cm}
\begin{figure*}[htbp]
\centering
\setlength{\tabcolsep}{0.2mm}
\begin{tabular}{ccm{\imagewidth}m{\imagewidth}m{\imagewidth}m{\imagewidth}m{\imagewidth}}
  \multirow{2}{*} &
  \rotatebox[origin=c]{90}{} &
  \epsfig{width=\imagewidth,file=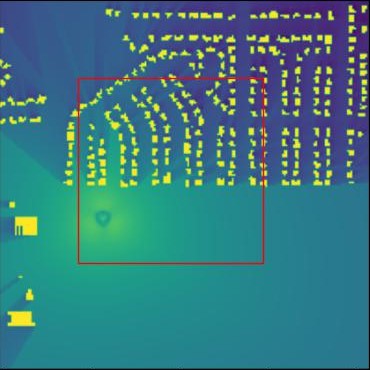} &
  \epsfig{width=\imagewidth,file=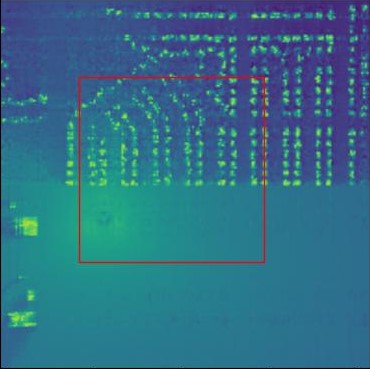} &
  \epsfig{width=\imagewidth,file=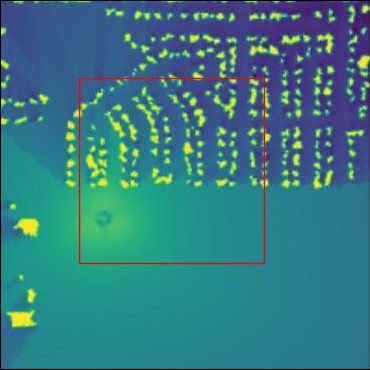} &
  \epsfig{width=\imagewidth,file=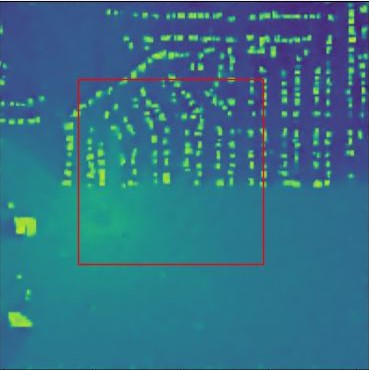} &
  \epsfig{width=\imagewidth,file=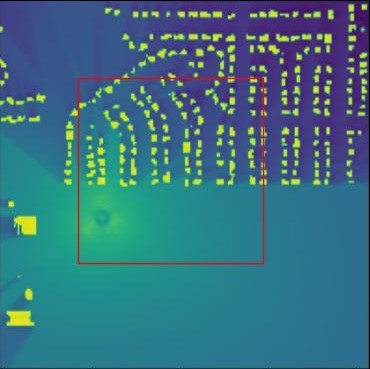} \\[0.1mm]
  &
  \rotatebox[origin=c]{90}{} &
  \epsfig{width=\imagewidth,file=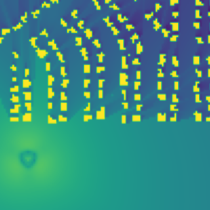} &
  \epsfig{width=\imagewidth,file=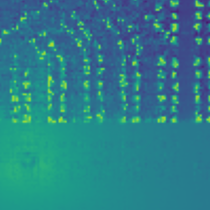} &
  \epsfig{width=\imagewidth,file=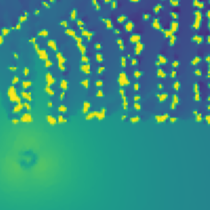} &
  \epsfig{width=\imagewidth,file=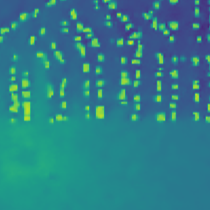} &
  \epsfig{width=\imagewidth,file=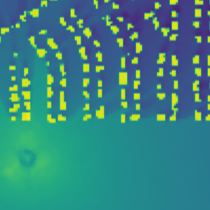} \\[0.5mm]
  & &   \multicolumn{1}{c}{\footnotesize{Ground-Truth}}   & \multicolumn{1}{c}{\footnotesize{HaLRTC}}  & \multicolumn{1}{c}{\footnotesize{RBF}}&
   \multicolumn{1}{c}{\footnotesize{FISTA-Net}} 
   & \multicolumn{1}{c}{\footnotesize{DULRTC-RME}} \\  [0.1mm]
\end{tabular}
\caption{\label{fig:Comparision} Visualization results of different methods; Detailed subimages in red squares are enlarged.}
\end{figure*}
\label{sec:experiments}
We use the BART-Lab Radiomap Dataset\footnote{https://github.com/BRATLab-UCD/Radiomap-Data}, which contains 2000 coarse-resolution radio maps across 5 frequency bands (1750MHz to 5750MHz). For simplicity, we selected maps at $\{2750, 3750, 4750\}$ MHz, and randomly cropped blocks near the base station of size $256 \times 256$, yielding 2000 tensors of size $256 \times 256 \times 3$ for training and testing.

To evaluate our proposed DULRTC-RME, we compare it with the representative tensor completion method HaLRTC \cite{b10}, RBF-based interpolation \cite{b4}, DL-based RadioUNet\cite{b5} and advanced unrolled-network-based FISTA-Net \cite{b12}. 
Among them, HaLRTC is a classical algorithm for tensor completion, which is widely used in the task of missing value completion of multidimensional data. 

RBF estimates unknown PSD values using a distance-based radial function.
RadioUNet is an efficient deep learning method to estimate the propagation path loss between transmitters and receivers while FISTA-Net integrates FISTA with algorithm unrolling to reconstruct video frames from compressed measurements for snapshot compressed imaging.

We train deep learning-based algorithms on 1200 dataset sets, using the remaining 800 for testing. Training is completed with the Adam optimizer, a batch size of 1, 25 epochs, and $K = 10$ unrolling iterations.

For quantitative evaluation, we fix the observation percentage at $10\%$ and assess reconstruction fidelity using PSNR and RMSE. The first two columns of Table~\ref{tab1} show that DULRTC-RME outperforms baselines, with higher PSNR and lower RMSE, indicating closer alignment with ground truth. This improvement is due to leveraging the low-rank properties of radio maps in both spatial and frequency domains and error-correcting regularisers.

\begin{table}[htbp]
\renewcommand{\arraystretch}{1.2}
\setlength{\tabcolsep}{8pt}
\vspace{-1em}
\centering
 \setlength{\tabcolsep}{2mm}
 \caption{Quantitative comparison of performance}
    \begin{tabular}{c|cccccc}
    \hline
    \hline  
          & PSNR$\uparrow$ & RMSE$\downarrow$ & Outage Error$\downarrow$  \bigstrut[t]\\
    \hline
    HaLRTC \cite{b10} & 21.54 & 0.1041  & 0.2415   \bigstrut[t]\\
    RBF \cite{b4} & 21.50 & 0.1055  & 0.2230   \\
    FISTA-Net \cite{b12} & 22.41 & 0.1154 & 0.2358  \bigstrut[t]\\
    RadioUNet \cite{b6} & 23.36 & 0.0905 & 0.2873 \bigstrut[t]\\
    DULRTC-RME & \textbf{25.81} & \textbf{0.0620} & \textbf{0.1261}
    \bigstrut[b]\\
    \hline
    \end{tabular}%
  \label{tab1}%
\vspace{0em}
\end{table}
\vspace{-2mm}

In applications like outage detection, a rough PSD distribution is often sufficient instead of an exact radio map. To assess the performance of DULRTC-RME in such cases, we conduct outage fault diagnosis referring to \cite{b8}. We define a location experiencing an outage if its PSD value falls below a set threshold. By converting the radio map into a binary outage map with normal and outage regions, we evaluate the error between the reconstructed and actual outage maps using an exemplary threshold. As shown in the third column of Table~\ref{tab1}, the DULRTC-RME algorithm achieves more accurate results, demonstrating its effectiveness for outage detection.

We also provide visual results of the reconstructed radio maps. Fig.~\ref{fig:Comparision} shows a comparison on a test image. Interpolation-based RBF and tensor completion-based HaLRTC fail to capture the underlying structure in PSD values, leading to homogenous and blurry reconstructions. Compared to FISTA-Net, DULRTC-RME produces more accurate radio maps by strictly enforcing the low-rank constraint and using learned regularizers for error compensation.

\begin{figure}[htbp]
	\centering
    \setlength{\tabcolsep}{0.5mm}
	\begin{minipage}{0.49\linewidth}
		\centering
		\includegraphics[width=1.02\linewidth]{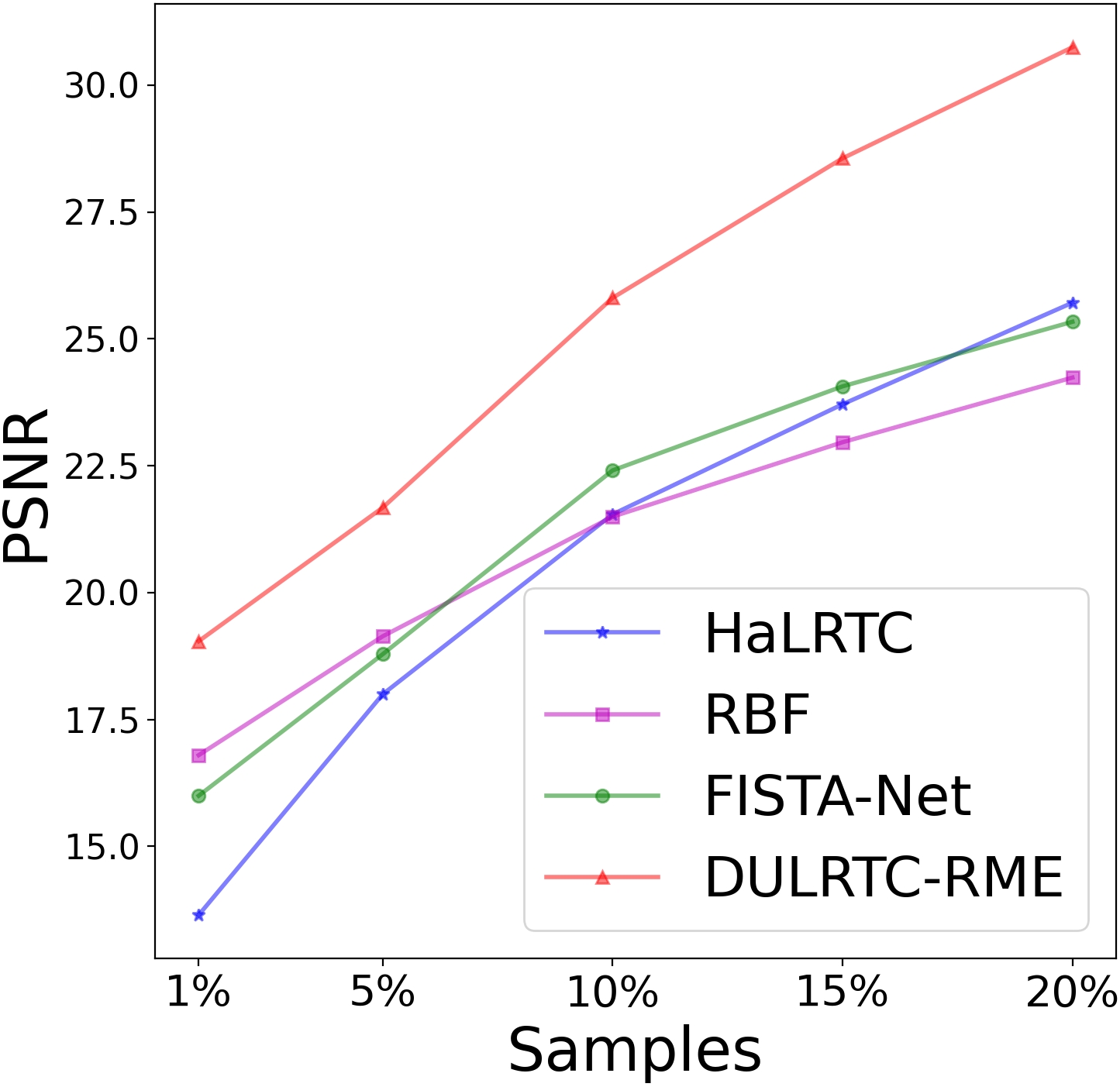}
	\end{minipage}
	\begin{minipage}{0.49\linewidth}
		\centering
		\includegraphics[width=1.02\linewidth]{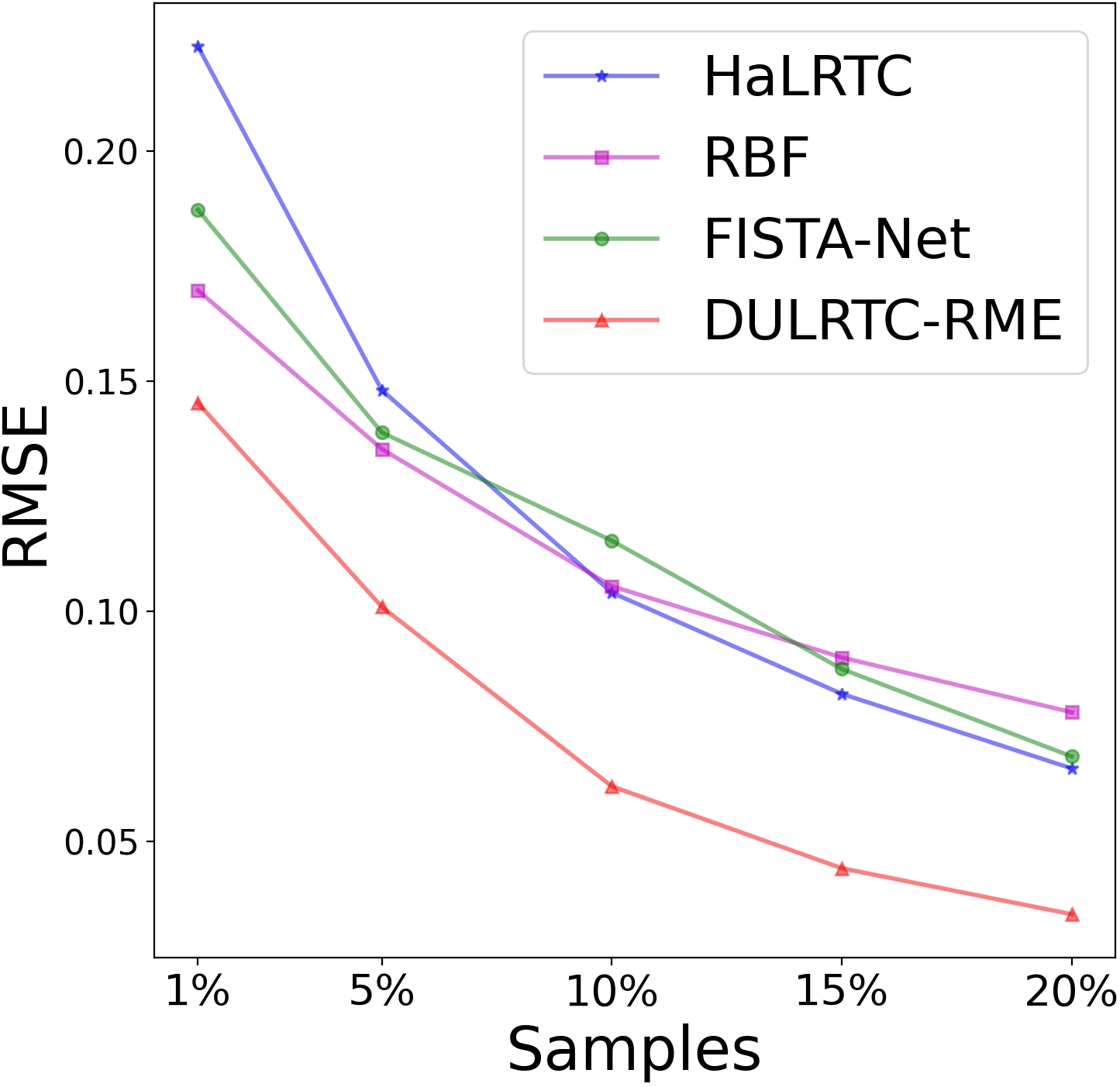}
	\end{minipage}
\caption{Performance for different sparsity levels.}
\label{fig:sparsity}
\end{figure}

In Fig.~\ref{fig:sparsity}, all methods are compared in different sparsity levels of measurements. The results show that DULRTC-RME consistently outperforms three baseline methods in PSNR and RMSE metrics. Notably, even with a minimal sparsity level of observations (1\%), DULRTC-RME still achieves relatively strong reconstruction performance.

\section{CONCLUSIONS}
\label{sec:conclusion}
By unrolling the low-rank tensor completion algorithm, we develop DULRTC-RME for radio map estimation from sparse observations, combining model-based and learning-based approaches. Specifically, spatial large fading and inter-frequency correlations of radio transmission are characterized as low-rank properties. The radio map estimation from sparse observations is formulated as a low-rank tensor completion optimization problem with learnable regularizers to capture structural information. To reduce the computation complexity and adaptively learn structural radio information, the ADMM iterative optimization process is completed by unrolling subproblems into modules of a DNN. In the proposed DULRTC-RME, closed-form solutions and learnable regularised terms are updated iteratively by training the network. Experimental results show that DULRTC-RME outperforms existing algorithms in reconstruction accuracy.

\vfill\pagebreak

\vspace{12pt}
\color{red}
\end{document}